\documentclass[
  aps,
  prb,
  reprint,
  superscriptaddress,
  amsmath,
  amssymb,
  nofootinbib,
  floatfix
]{revtex4-2}

\usepackage{graphicx}
\usepackage{bm}
\usepackage{booktabs}
\usepackage{siunitx}
\usepackage{xcolor}
\usepackage{textcomp}
\usepackage[percent]{overpic}
\usepackage{subcaption}
\usepackage[compatibility=false]{caption}
\DeclareCaptionJustification{revtexflush}{\leftskip0pt\rightskip0pt\parfillskip0pt plus1fil}
\usepackage{hyperref}

\hypersetup{
  colorlinks=true,
  linkcolor=blue,
  citecolor=blue,
  urlcolor=blue
}

\newcommand{\DirectGap}{0.54}
\newcommand{\IndirectGap}{0.24}

\begin{document}

\title{Giant Bulk-Rashba Splitting in Polar Topological Insulator BiSbTeSe\texorpdfstring{$_2$}{2}}

\author{Ritam Chakraborty}
\email{ritamchakraborty454@gmail.com}
\affiliation{Theoretical Sciences Unit, Jawaharlal Nehru Centre for Advanced Scientific Research, Bangalore 560064, India}

\date{July 29, 2026}

\begin{abstract}
Bulk-Rashba spin splitting is forbidden in tetradymite topological insulators like Bi$_2$Se$_3$ or Bi$_2$Te$_3$, since their quintuple-layer stacking preserves inversion symmetry. We show that BiSbTeSe$_2$ escapes this restriction: in the Se-Bi-Se-Sb-Te sequence, the structure loses its inversion center, reducing the point group symmetry at $\Gamma$ from $D_{3d}$ to $C_{3v}$. First-principles density functional calculations with spin-orbit coupling show that this ordered structure retains bulk band inversion and a linearly dispersive surface state of a strong topological insulator. Additionally, its bulk bands acquire a pronounced linear-in-$k$ spin splitting away from $\Gamma$. 
Fitting the conduction- and valence-band doublets to symmetry-constrained two-band $k\cdot p$ Hamiltonians, we extract intrinsic linear Rashba coefficients of $\alpha_{\mathrm{CB}}\approx2.66~\mathrm{eV\,\text{\AA}}$ and
$\alpha_{\mathrm{VB}}\approx0.35~\mathrm{eV\,\text{\AA}}$. The conduction-band value places ordered BiSbTeSe$_2$ among the strongest bulk-Rashba topological-insulator systems reported to date and approaches the coupling found in the benchmark polar Rashba semiconductor BiTeI.
 Sublattice ordering thus provides a route to giant bulk
spin--momentum locking that coexists with protected
topological surface states, offering a platform in which
bulk Rashba and topological surface contributions to
spin and charge transport can be investigated within the
same material.
\end{abstract}

\maketitle

\section{Introduction}

Three-dimensional topological insulators host metallic surface states whose spin and momentum are locked by time-reversal symmetry, suppressing elastic backscattering from nonmagnetic disorder~\cite{Hasan2010,Qi2011}. The Bi$_2$Se$_3$ family provides the canonical materials platform for this physics, with a single Dirac cone in the surface Brillouin zone~\cite{Zhang2009,Xia2009,Chen2009}. In practice, however, unintentional bulk carriers often obscure the surface response of Bi$_2$Se$_3$ and Bi$_2$Te$_3$.

Alloying across the Bi$_{2-x}$Sb$_x$Te$_{3-y}$Se$_y$ series has therefore been used to suppress bulk conduction. The quaternary compound BiSbTeSe$_2$ (BSTS) is among the most bulk-insulating tetradymites~\cite{Ren2010,Ando2013}. Photoemission and density functional calculations have identified a band-inverted electronic structure and a topological surface state in BSTS-related compositions~\cite{Lohani2017}. Transport measurements also reveal coexisting bulk and surface channels over experimentally relevant temperature ranges ~\cite{Shyaga2025}.

The stacking of atomic layers in BSTS is central to its electronic symmetry. The four distinct atomic species can be arranged in several inequivalent sequences within a quintuple layer. Raman spectroscopy and first-principles calculations identify the Se--Bi--Se--Sb--Te sequence as energetically favorable~\cite{German2019}. Unlike the inversion-symmetric Te--Bi--Te--Bi--Te sequence of Bi$_2$Te$_3$, this ordered layer is chemically inequivalent on its two sides. Consequently, inversion symmetry is broken within the primitive cell while the threefold rotation and vertical mirror symmetries are retained.

The combination of broken inversion symmetry and an inverted band structure provides a route to realizing bulk spin-split states and topological surface states within the same material. BiTeI is the best-known example of a polar semiconductor with giant bulk-Rashba splitting at ambient pressure~\cite{Ishizaka2011}. Under hydrostatic pressure, it is predicted to undergo a band inversion and enter a strong topological-insulator phase while retaining its noncentrosymmetric structure and large Rashba coupling~\cite{Bahramy2012}. The related compound BiTeCl hosts a single topological Dirac fermion at ambient conditions as a consequence of its strong structural inversion asymmetry~\cite{Chen2013}, while experiments on BiTeI have separately resolved the bulk and surface contributions to the observed Rashba splitting~\cite{Landolt2012}. Similar functionality has also been proposed in BiTeI/Bi$_2$Te$_3$ heterostructures, where a giant-Rashba-split layer is coupled to a topological surface state~\cite{Zhou2014}. Ferroelectric Rashba semiconductors such as GeTe and SnTe provide another class of examples~\cite{DiSante2013, Liebmann2016}, in which strain, alloying, or ferroelectric distortions can drive transitions between Rashba-split, topological-crystalline-insulator, and $\mathbb{Z}_2$ topological-insulator phases~\cite{Plekhanov2014, Hsieh2012}.

These studies establish that bulk-Rashba splitting and nontrivial topology are compatible. Such coexistence is particularly attractive for spintronic applications because it combines bulk Rashba-driven spin--charge conversion and current-induced spin polarization with a topological surface channel, while offering the possibility of tuning the relative bulk and surface contributions through the chemical potential. In most of the known examples, however, their coexistence is obtained through pressure, strain, electric fields, heterostructure engineering, or the selection of a comparatively specialized compound. It is therefore important to determine whether both properties can arise intrinsically, at ambient conditions, in an experimentally established bulk-insulating topological insulator.

BiSbTeSe$_2$ is particularly well suited to address this question. It is among the most bulk-insulating members of the tetradymite family~\cite{Ren2010, Ando2013} and is already widely used to investigate surface-dominated and multichannel transport~\cite{Shyaga2025}. At the same time, its energetically preferred Se--Bi--Se--Sb--Te ordering is intrinsically polar because the two sides of the quintuple layer are chemically inequivalent~\cite{German2019}. The central question is therefore whether this chemically ordered bulk phase retains its strong topological character while also developing a sizeable intrinsic Rashba splitting. A quantitative description further requires separating the true $k\rightarrow 0$ Rashba coefficient from finite-momentum slopes and determining whether higher-order terms allowed by the $C_{3v}$ symmetry are significant. Establishing these properties would identify chemical sublattice ordering as a direct route for introducing bulk spin functionality into an already important topological-insulator platform.

Here, we address these questions using fully relativistic density functional theory based calculations, momentum-resolved spin textures, symmetry-constrained two-band $k\cdot p$ models, and Wannier-based topological diagnostics. Our calculations establish ordered BiSbTeSe$_2$ in $R3m$ phase as a noncentrosymmetric strong topological insulator with pronounced bulk-Rashba splitting. For the conduction-band doublet, we obtain an intrinsic linear Rashba coefficient of
$\alpha_{\mathrm{CB}}=2.666\pm0.005~\mathrm{eV\,\text{\AA}}$, together with a sizeable negative cubic radial correction. The more weakly split valence-band doublet has
$\alpha_{\mathrm{VB}}=0.345\pm0.005~\mathrm{eV\,\text{\AA}}$. Wilson-loop evolution and the semi-infinite surface spectral function together establish that the same ordered structure remains in the strong topological phase. These results demonstrate that sublattice ordering in BiSbTeSe$_2$ produces strong bulk spin--momentum locking without eliminating its topological surface channel.

\section{Computational details}
\label{sec:methods}

All calculations were performed within nonmagnetic density functional theory using the \textsc{Quantum ESPRESSO} package~\cite{Giannozzi2009,Giannozzi2017}. Interactions between the ionic cores and valence electrons were described using ultrasoft pseudopotentials~\cite{Vanderbilt1990}; spin--orbit coupling was included through fully relativistic pseudopotentials~\cite{DalCorso2014}. The exchange-correlation potential was treated using the Perdew--Burke--Ernzerhof (PBE) form of Generalized-Gradient Approximation (GGA)~\cite{PBE1996}. The van-der-Waals dispersion correction was incorporated  using the semi-empirical DFT-D3 formalism~\cite{Grimme2010}. The Kohn--Sham wavefunctions and charge densities were
expanded in plane-wave basis sets with kinetic-energy
cutoffs of 50 and 500 Ry, respectively. Lattice vectors and internal coordinates were fully relaxed using the Broyden--Fletcher--Goldfarb--Shanno algorithm~\cite{Broyden1973,Fletcher1970,Goldfarb1970,Shanno1970} until every force component was below $10^{-3}$~Ry/Bohr. The primitive Brillouin zone was sampled using an $18\times18\times9$ Monkhorst--Pack mesh~\cite{Monkhorst1976} for structural relaxation.

 The calculations employed an oblique five-atom primitive cell containing one quintuple layer of the rhombohedral R3m structure. The vectors $\mathbf{a}_1$ and $\mathbf{a}_2$ span the hexagonal basal plane, while $\mathbf{a}_3$ is an oblique primitive translation that includes the lateral shift between neighboring quintuple layers (see Fig.~\ref{fig:BSTS_crystal_structure}). The crystallographic threefold axis ($C_3$) is not parallel to $\mathbf{a}_3$; it is normal to the $\mathbf{a}_1$--$\mathbf{a}_2$ plane.
Symmetry analysis was performed using \textsc{pymatgen}~\cite{Ong2013}. The relaxed ordered structure belongs to the noncentrosymmetric space group $R3m$ (No.~160), with point group $3m\equiv C_{3v}$.

Maximally localized Wannier functions were constructed using \textsc{Wannier90}~\cite{Marzari1997,Souza2001,Mostofi2008,Mostofi2014}. The fully relativistic Kohn--Sham states were projected onto the $p$ orbitals of Bi, Sb, Te, and Se, producing a 30-band spinor Wannier Hamiltonian. The topological indices and semi-infinite surface spectral function were evaluated using \textsc{WannierTools}~\cite{Wu2018}. The $\mathbb{Z}_2$ indices were obtained from Wilson-loop evolution of the hybrid Wannier charge centers~\cite{FuKane2006,Yu2011,Soluyanov2011}, and the surface Green function was calculated using the iterative method of Sancho \emph{et al.}~\cite{Sancho1985}. Details of the Rashba analysis are provided in the Supplemental Material.

\section{Results and discussion}

\subsection{Crystal structure and band inversion}
\label{sec:structure}

\begin{figure*}[!ht]
    \centering
    \includegraphics[width=0.90\textwidth]{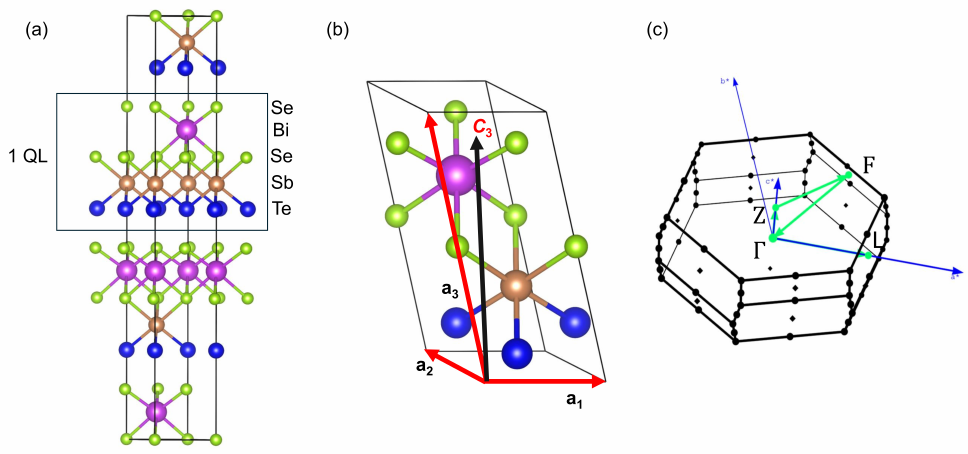}
    \caption{
Crystal structure and bulk Brillouin zone of ordered BiSbTeSe$_2$ in the Se--Bi--Se--Sb--Te stacking configuration. 
(a) Conventional hexagonal cell containing three quintuple layers (QLs); the boxed region marks one QL and indicates its atomic-layer sequence. 
(b) Oblique five-atom primitive cell used in all the first-principles calculations performed in this paper, with lattice vectors $\mathbf{a}_1$, $\mathbf{a}_2$, and $\mathbf{a}_3$. The crystallographic threefold axis $C_3$ is normal to the hexagonal basal plane spanned by $\mathbf{a}_1$ and $\mathbf{a}_2$ and is not parallel to the oblique vector $\mathbf{a}_3$. 
(c) Corresponding primitive bulk Brillouin zone, showing the high-symmetry points $\Gamma$, $Z$, $F$, and $L$ and the path $\Gamma$--$Z$--$F$--$\Gamma$--$L$ used for the bulk band-structure calculation. Bi, Sb, Te, and Se atoms are shown in magenta, orange, blue, and green, respectively.
}
\label{fig:BSTS_crystal_structure}
\end{figure*}

\begin{figure*}[!ht]
    \centering
\includegraphics[width=0.90\textwidth]{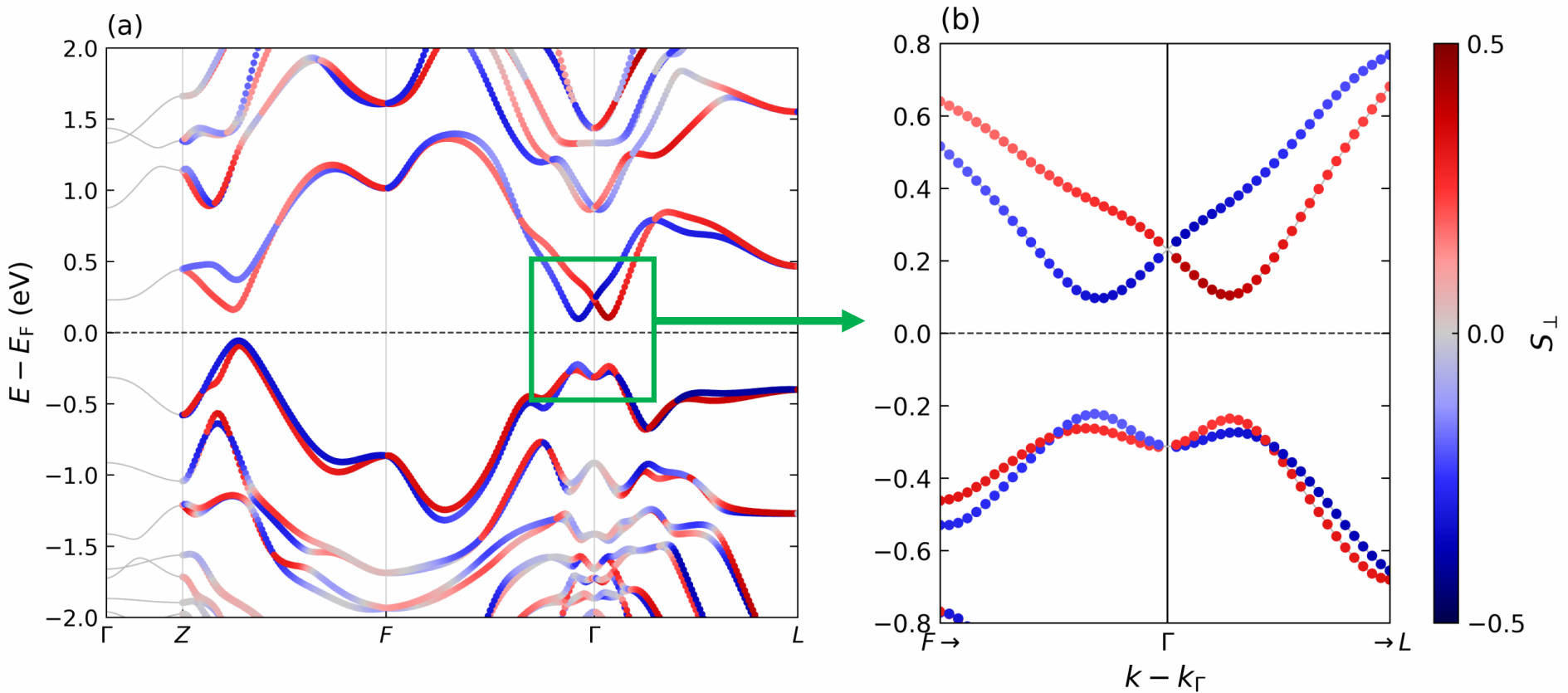}    
\caption{
Spin-resolved electronic band structure of ordered BiSbTeSe$_2$ calculated with spin--orbit coupling.
(a) Bulk band structure along the high-symmetry path
$\Gamma$--$Z$--$F$--$\Gamma$--$L$, colored by the signed transverse
in-plane spin component
$S_{\perp}=-t_y\langle S_x\rangle+t_x\langle S_y\rangle$, where
$(t_x,t_y)$ is the local in-plane unit tangent to the momentum path.
Positive and negative values of $S_{\perp}$ therefore represent opposite
spin orientations transverse to the direction of propagation.
(b) Enlarged view around the second $\Gamma$ point along
$F$--$\Gamma$--$L$. The two branches meet at $\Gamma$, as required by
Kramers degeneracy, and split progressively away from $\Gamma$.
The opposite signs of $S_{\perp}$ on the split branches demonstrate
their opposite spin helicities and provide a clear signature of bulk
Rashba spin--momentum locking. Energies are shown relative to the Fermi
level.}
\label{fig:spin-colored-band-str}
\end{figure*}

BSTS can be represented either by an oblique five-atom primitive cell containing one quintuple layer or by a conventional hexagonal cell containing three quintuple layers. We compared the Se--Bi--Se--Sb--Te, Se--Sb--Se--Bi--Te, and Se--Sb--Bi--Te--Se layer sequences. The Se--Bi--Se--Sb--Te configuration has the lowest total energy and was therefore used in all subsequent calculations.

The chemical inequivalence of the two sides of the ordered quintuple layer removes inversion symmetry but preserves the threefold rotation and vertical mirror operations, yielding the polar space group $R3m$. Fig.~\ref{fig:BSTS_crystal_structure}(a) and (b) show three-quintuple-layers thick conventional hexagonal cell and one-quintuple-layer thick primitive representation respectively, where the $\mathbf{a}_3$ is at an angle with respect to the crystallographic $C_3$ axis. The primitive Brillouin zone is shown in Fig.~\ref{fig:BSTS_crystal_structure}(c). The near $\Gamma$ Rashba analysis was done in the $\mathbf{a}_1 - \mathbf{a}_2$ ($\mathbf{b}_1 - \mathbf{b}_2$) plane, perpendicular to $C_3$-axis.

The electronic band structure calculated incorporating SOC is shown in Fig.~\ref{fig:spin-colored-band-str}. The direct gap at $\Gamma$ is approximately $\DirectGap$~eV and the indirect band gap is approximately $\IndirectGap$~eV.  Orbital projections reveal an inversion between cation and anion-derived $p$ states close to $\Gamma$ (see Fig.~\ref{fig:S1}); the topmost valence states acquire predominantly cationic character, while the lowest conduction states acquire predominantly anionic character. Because the structure lacks inversion symmetry, this orbital inversion is suggestive but is not by itself a complete topological classification; the latter is established below from Wilson loops and the surface spectrum.

\subsection{Spin-momentum locking and bulk-Rashba spin textures}
\label{sec:spin_texture}

Figure~\ref{fig:spin-colored-band-str} shows the spin-resolved bulk band structure of
ordered BiSbTeSe$_2$, calculated with spin--orbit coupling. The bands are
colored by the signed transverse spin component
$S_\perp = -t_y\langle S_x\rangle + t_x\langle S_y\rangle$, where
$(t_x,t_y)$ is the local tangent to the momentum path, so the two colors
correspond to spin orientations pointing in opposite senses perpendicular
to the electron momentum. The full path in the left panel of 
Fig.~\ref{fig:spin-colored-band-str} shows pronounced spin polarization on several
bands, and the enlarged view around $\Gamma$ in the right panel of 
Fig.~\ref{fig:spin-colored-band-str} resolves the band-edge splitting directly.
The branches of each doublet remain degenerate at $\Gamma$, as required
by time-reversal symmetry, but separate immediately away from $\Gamma$
with opposite sign of $S_\perp$; the signature of classic Rashba splitting of the  spin--momentum locked branches. So, at $\Gamma$, we have two sets of Kramers doublet, one corresponding to valence band and the other corresponding to conduction band.

\begin{figure*}[!ht]
    \centering
    \begin{minipage}[b]{0.48\textwidth}
        \centering
        \begin{overpic}[width=\textwidth]{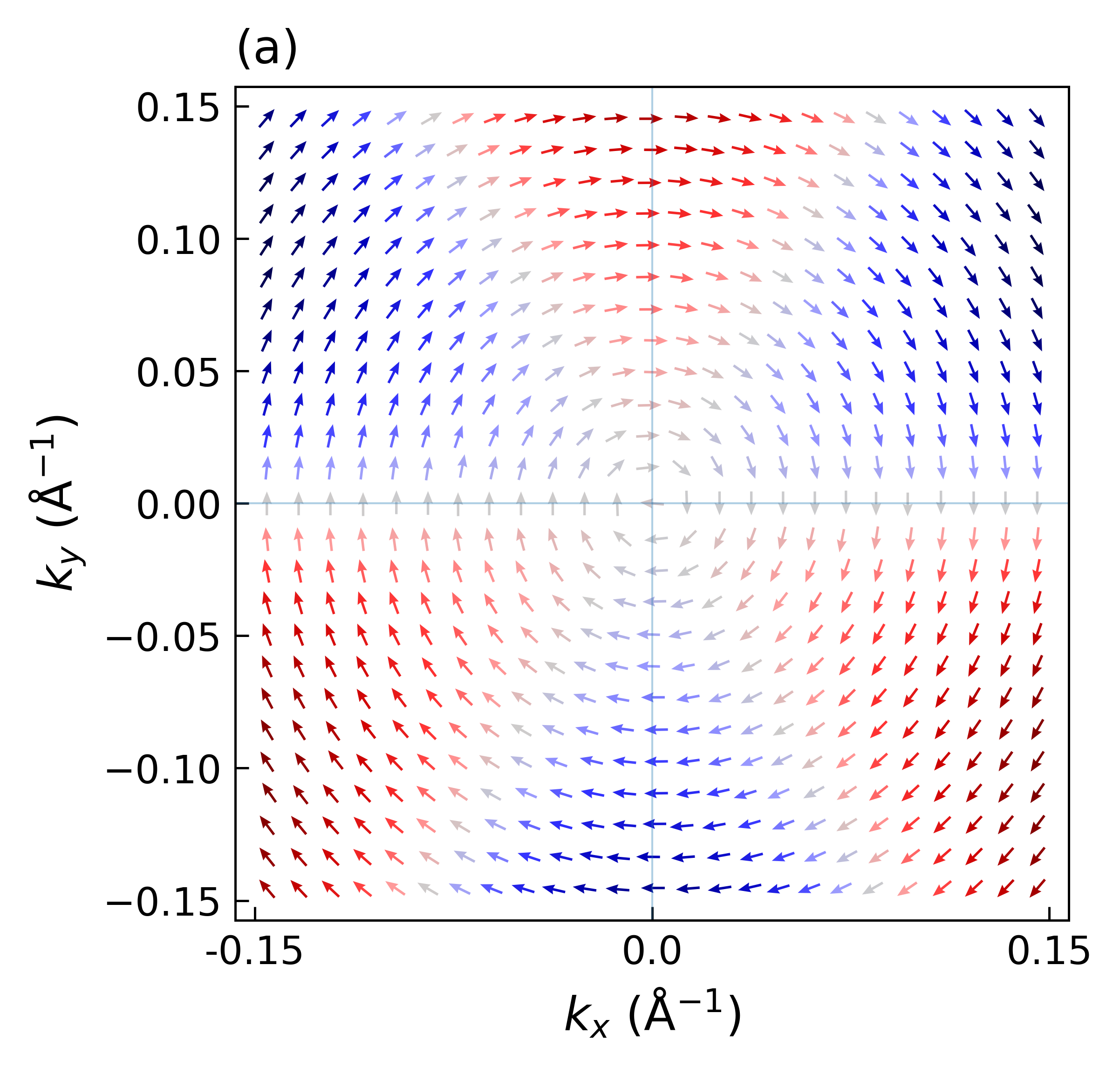}
        \end{overpic}
    \end{minipage}
    \hfill
    \begin{minipage}[b]{0.48\textwidth}
        \centering
        \begin{overpic}[width=\textwidth]{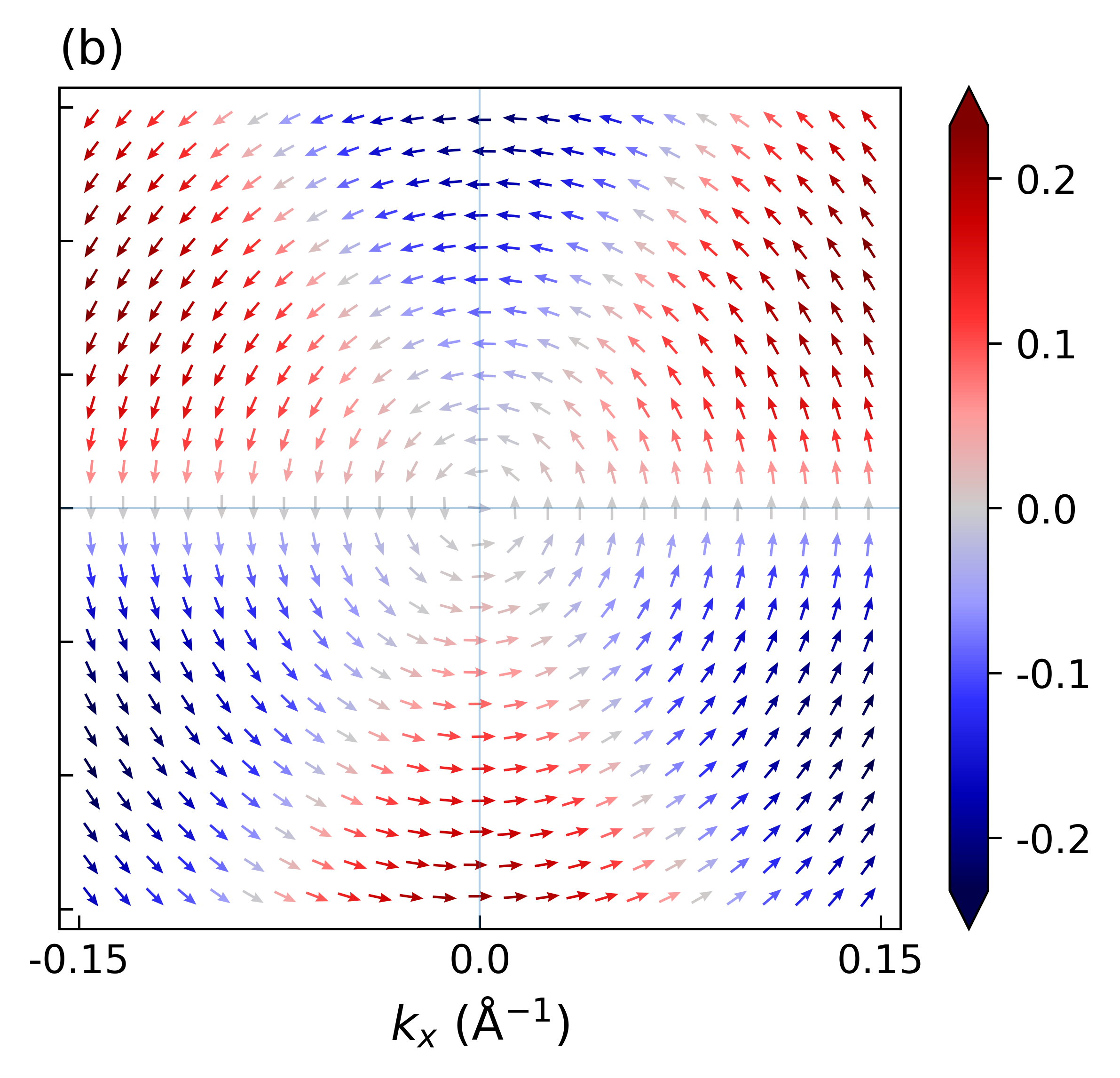} 
        \end{overpic}
    \end{minipage}
    
\caption{Calculated physical-spin texture of the two Rashba-split conduction-band 
branches in the momentum plane perpendicular to the crystallographic 
$C_3$ axis, $\mathbf{k}\cdot\hat{\mathbf{z}}_{C_3}=0$. The arrows show 
the in-plane components 
$(\langle\sigma_x\rangle,\langle\sigma_y\rangle)$, with their lengths 
proportional to the in-plane spin magnitude, while the color scale 
represents the component $\langle\sigma_z\rangle$ parallel to the 
$C_3$ axis. The corresponding physical spin is 
$\langle\mathbf{S}\rangle=(\hbar/2)\langle\boldsymbol{\sigma}\rangle$. 
Panels (a) and (b) correspond to the lower and upper energy-ordered 
branches (CB and CB+1), respectively. The approximately transverse in-plane spin 
orientation demonstrates Rashba-type spin--momentum locking, while the 
alternating angular pattern of the $C_3$-parallel component is consistent 
with the cubic $C_{3v}$ spin--orbit term.}
\label{fig:cb_spin_texture}
\end{figure*}

\begin{figure*}[!ht]
    \centering
    \begin{minipage}[b]{0.48\textwidth}
        \centering
        \begin{overpic}[width=\textwidth]{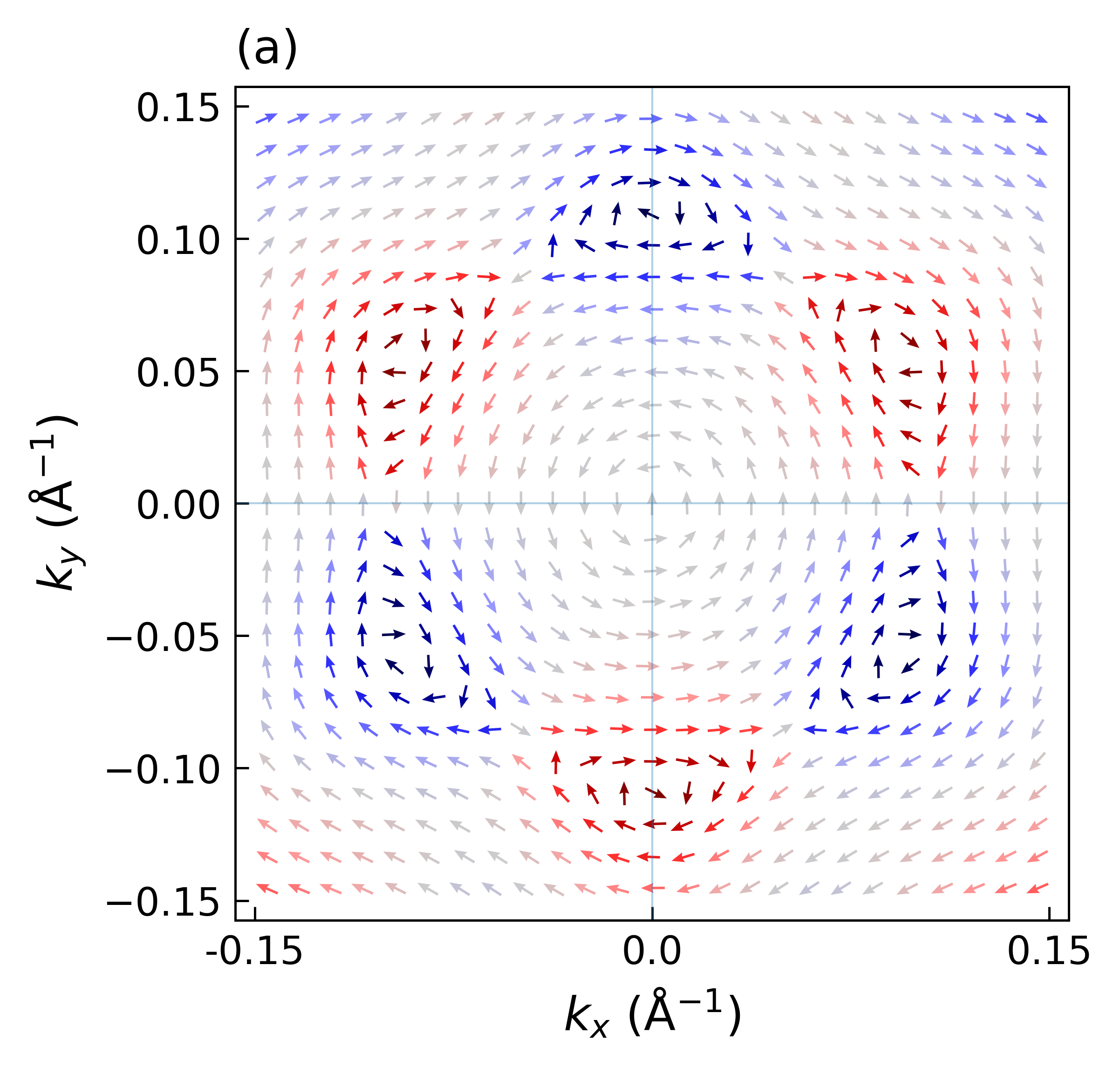}
        \end{overpic}
    \end{minipage}
    \hfill
    \begin{minipage}[b]{0.48\textwidth}
        \centering
        \begin{overpic}[width=\textwidth]{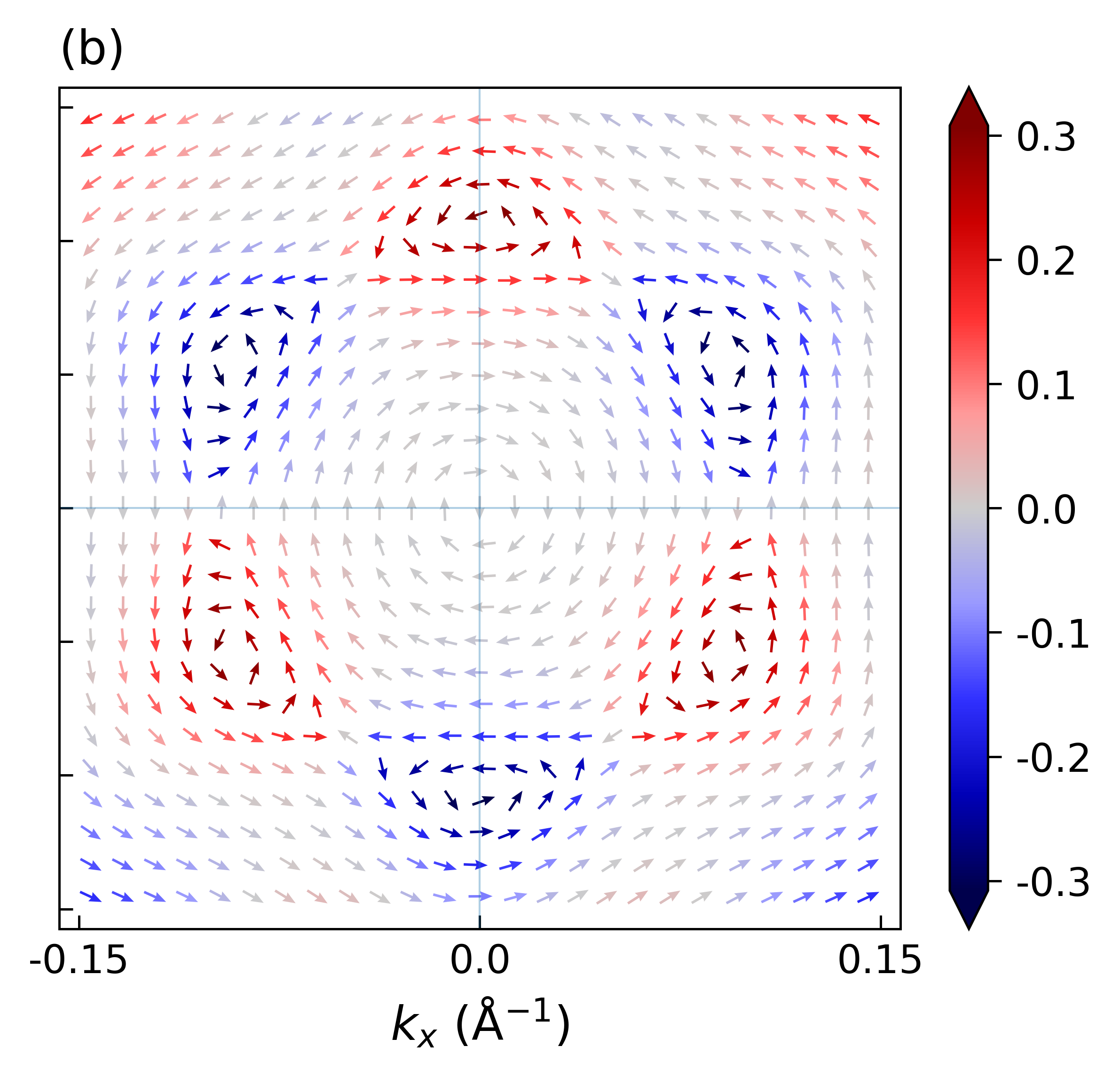} 
        \end{overpic}
    \end{minipage}
    
\caption{Calculated physical-spin texture of the two Rashba-split valence-band 
branches in the momentum plane perpendicular to the crystallographic 
$C_3$ axis, $\mathbf{k}\cdot\hat{\mathbf{z}}_{C_3}=0$. The arrows show 
the in-plane components 
$(\langle\sigma_x\rangle,\langle\sigma_y\rangle)$, with their lengths 
proportional to the in-plane spin magnitude, while the color scale 
represents the component $\langle\sigma_z\rangle$ parallel to the 
$C_3$ axis.  
Panels (a) and (b) correspond to the lower and upper energy-ordered 
branches (VB-1 and VB), respectively. The approximately transverse in-plane spin 
orientation demonstrates Rashba-type spin--momentum locking, while the 
alternating angular pattern of the $C_3$-parallel component is consistent 
with the cubic $C_{3v}$ spin--orbit term.}
\label{fig:vb_spin_texture}
\end{figure*}

A complementary, momentum-resolved view is given by the spin textures of
Figs.~\ref{fig:cb_spin_texture} and \ref{fig:vb_spin_texture}, which show the spin expectation values in the momentum plane perpendicular to the
crystallographic $C_3$ axis for conduction band and valence band doublets respectively. The in-plane spin components are
represented by arrows, while the component parallel to $C_3$ (out-of-plane) is
represented by the color scale.

\subsubsection{Conduction-band doublet}

At $\Gamma$, the two conduction-band states form a degenerate Kramers pair. Since any linear combination of the two degenerate states is also an allowed eigenstate, an individual branch helicity is not uniquely defined exactly at $\Gamma$. At finite in-plane momentum, however, the degeneracy is lifted and the lower and upper energy-ordered branches acquire opposite spin helicities. Close to $\Gamma$, the in-plane spin is almost entirely perpendicular to the momentum $\mathbf{k}$, which is the characteristic signature of Rashba spin--momentum locking. This behaviour is also reflected in Fig.~\ref{fig:spin-colored-band-str} by the opposite signs of the transverse spin component $S_{\perp}$ on the two split branches.

The spin component parallel to the crystallographic $C_{3}$ axis is smaller than the in-plane component and exhibits a pronounced angular modulation. At a fixed nonzero momentum magnitude, this component varies approximately as $\sin 3\phi$, where $\phi$ is the in-plane polar angle. It therefore vanishes along radial directions separated by $60^{\circ}$ and reverses sign between adjacent angular sectors, producing six alternating positive and negative lobes around $\Gamma$. The sign pattern is reversed between the two energy-ordered branches. This threefold angular modulation is the spin-texture signature of the cubic $C_{3v}$ spin--orbit term introduced in Sec.~\ref{sec:kp_model}.

Since the system is nonmagnetic, time-reversal symmetry requires
\begin{equation}
E_{n}(\mathbf{k}) = E_{n}(-\mathbf{k}),
\qquad
\left\langle \mathbf{S} \right\rangle_{n}(-\mathbf{k})
=
-\left\langle \mathbf{S} \right\rangle_{n}(\mathbf{k}).
\end{equation}
Thus, the states belonging to the same energy-ordered branch at $\mathbf{k}$ and $-\mathbf{k}$ are Kramers partners. By contrast, the lower- and upper-energy states at the same finite momentum are not Kramers partners, even though they carry approximately opposite spin helicities.

\subsubsection{Valence-band doublet}

The two valence-band branches display the same qualitative signatures of Rashba spin--momentum locking as the conduction-band  (see Fig.~\ref{fig:vb_spin_texture}). Close to $\Gamma$, their in-plane spin components wind approximately tangentially around the momentum-space origin, with opposite helicities for the two energy-ordered branches. The spin component parallel to the crystallographic $C_{3}$ axis also shows a pronounced threefold angular modulation, with its sign reversing between adjacent angular sectors and between the two branches.

The valence-band energy splitting is substantially smaller than that of the conduction-band doublet. Consequently, its quantitative description is more sensitive to the momentum range included in the fit and to higher-order corrections away from $\Gamma$. In this section, the calculated spin texture is therefore used only to establish the Rashba character of the valence-band doublet qualitatively. The symmetry-constrained Hamiltonian and the quantitative extraction of the corresponding model coefficients are introduced and discussed separately in Sec.~\ref{sec:kp_model}.

\subsection{Symmetry-constrained two-band \texorpdfstring{$k\cdot p$}{k.p} model and extraction of the Rashba parameters}
\label{sec:kp_model}

At $\Gamma$, the conduction-band edge and the valence-band edge each form a Kramers-degenerate doublet. We describe these two doublets independently using separate two-band Hamiltonians. The purpose of the model is to describe the local dispersion and spin splitting close to $\Gamma$; it is not intended to represent the complete band inversion or the global topological band structure.

For either doublet, denoted by
\[
D\in\{\mathrm{CB},\mathrm{VB}\},
\]
the Hamiltonian is written as
\begin{equation}
H_D(\mathbf{k})
=
\epsilon_D(\mathbf{k})\sigma_0
+
\mathbf{h}_D(\mathbf{k})\cdot\boldsymbol{\sigma},
\label{eq:general_two_band}
\end{equation}
where $\sigma_0$ is the $2\times2$ identity matrix and
$\boldsymbol{\sigma}=(\sigma_x,\sigma_y,\sigma_z)$ are Pauli matrices acting within the spin--orbital Kramers-doublet subspace. These Pauli matrices describe a model pseudospin and should not, in general, be identified quantitatively with the physical-spin operator obtained from the DFT wavefunctions.

The local Cartesian coordinates are chosen such that the $z$ axis is parallel to the crystallographic $C_3$ axis, while $k_x$ and $k_y$ lie in the plane perpendicular to $C_3$. The $x$ axis is chosen within one of the vertical mirror planes of the $C_{3v}$ point group. We define
\begin{equation}
k^2=k_x^2+k_y^2,
\qquad
k_x=k\cos\phi,
\qquad
k_y=k\sin\phi .
\label{eq:k_coordinates}
\end{equation}

Following the symmetry structure of the hexagonally warped $C_{3v}$ Hamiltonian introduced by Fu and its higher-order Rashba generalisation~\cite{Fu2009,Vajna2012}, we use
\begin{align}
H_D(\mathbf{k})
={}&
\epsilon_D(\mathbf{k})\sigma_0
+
\left(\alpha_D+\beta_D k^2\right)
\left(k_x\sigma_y-k_y\sigma_x\right)
\nonumber\\
&+
\lambda_D
\left(3k_x^2k_y-k_y^3\right)\sigma_z .
\label{eq:kp_hamiltonian}
\end{align}

The spin-independent part is represented by
\begin{equation}
\epsilon_D(\mathbf{k})
=
E_{0,D}+c_{2,D}k^2+c_{4,D}k^4 .
\label{eq:band_centre}
\end{equation}
It describes the average dispersion of the two branches and does not contribute to their energy separation.

The term proportional to $\alpha_D$ is the conventional linear Rashba term. It produces an in-plane effective field perpendicular to $\mathbf{k}$ and therefore gives two branches with opposite helicities. Infinitesimally close to $\Gamma$, the splitting generated by this term is linear in $k$.

The coefficient $\beta_D$ describes an isotropic cubic correction to the in-plane Rashba field. The corresponding momentum-dependent in-plane coefficient is
\begin{equation}
\alpha_D^{\mathrm{eff}}(k)
=
\alpha_D+\beta_D k^2 .
\label{eq:alpha_effective}
\end{equation}
Thus, $\alpha_D$ is the intrinsic $k\rightarrow0$ Rashba coefficient, whereas $\alpha_D^{\mathrm{eff}}(k)$ describes the effective coupling at finite momentum.

The term proportional to $\lambda_D$ describes the anisotropic cubic warping allowed by $C_{3v}$ symmetry. In the present coordinate convention,
\begin{equation}
w(\mathbf{k})
=
3k_x^2k_y-k_y^3
=
k^3\sin(3\phi).
\label{eq:warping_harmonic}
\end{equation}
This term produces a pseudospin component parallel to the $C_3$ axis. It vanishes along $\phi=0^\circ$ and reaches its maximum magnitude along $\phi=30^\circ$. The sine form used here is related to the more familiar $k^3\cos(3\phi)$ form by a $30^\circ$ rotation of the in-plane coordinate axes and does not represent a different physical interaction.

Writing Eq.~(\ref{eq:kp_hamiltonian}) as
\[
H_D(\mathbf{k})
=
\epsilon_D(\mathbf{k})\sigma_0
+
\mathbf{h}_D(\mathbf{k})\cdot\boldsymbol{\sigma},
\]
the effective pseudospin field is
\begin{equation}
\mathbf{h}_D(\mathbf{k})
=
\begin{pmatrix}
-\left(\alpha_D+\beta_Dk^2\right)k_y\\[2mm]
 \left(\alpha_D+\beta_Dk^2\right)k_x\\[2mm]
 \lambda_D\left(3k_x^2k_y-k_y^3\right)
\end{pmatrix}.
\label{eq:effective_field}
\end{equation}

The two eigenvalues are
\begin{equation}
E_{D,\pm}(\mathbf{k})
=
\epsilon_D(\mathbf{k})
\pm d_D(\mathbf{k}),
\label{eq:model_eigenvalues}
\end{equation}
where
\begin{align}
d_D(\mathbf{k})
=
\Big[
&
\left\{\left(\alpha_D+\beta_Dk^2\right)k\right\}^{2}
\nonumber\\
&+
\lambda_D^2
\left(3k_x^2k_y-k_y^3\right)^2
\Big]^{1/2}.
\label{eq:model_half_splitting}
\end{align}
The full separation between the two branches is therefore
\[
E_{D,+}-E_{D,-}=2d_D.
\]
In the $k\rightarrow0$ limit,
\begin{equation}
d_D(k)\simeq |\alpha_D|k,
\qquad
E_{D,+}-E_{D,-}\simeq2|\alpha_D|k.
\label{eq:linear_limit}
\end{equation}

\begin{figure*}[t]
    \centering
    \begin{subfigure}[t]{0.48\textwidth}
        \centering
        \includegraphics[width=\linewidth]{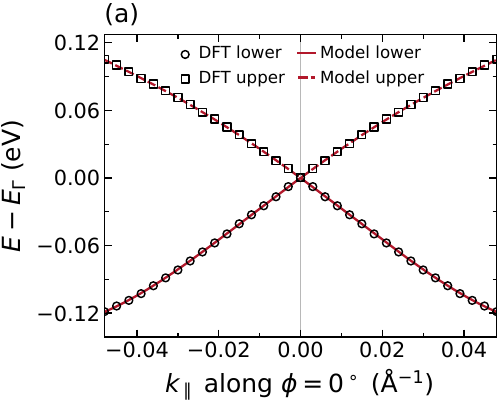}
        \label{fig:panel_a}
    \end{subfigure}
    \hfill
    \begin{subfigure}[t]{0.48\textwidth}
        \centering
         \includegraphics[width=\linewidth]{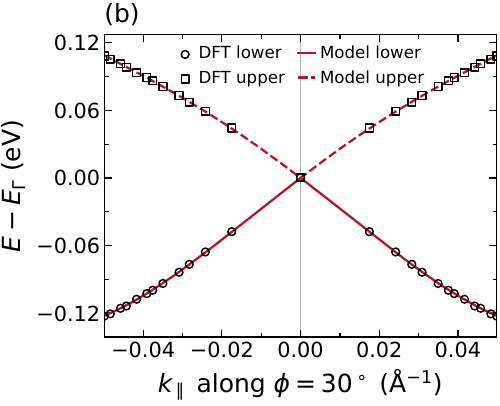}
        \label{fig:panel_b}
    \end{subfigure}

    \vspace{0.15in}

    \begin{subfigure}[t]{0.48\textwidth}
        \centering
         \includegraphics[width=\linewidth]{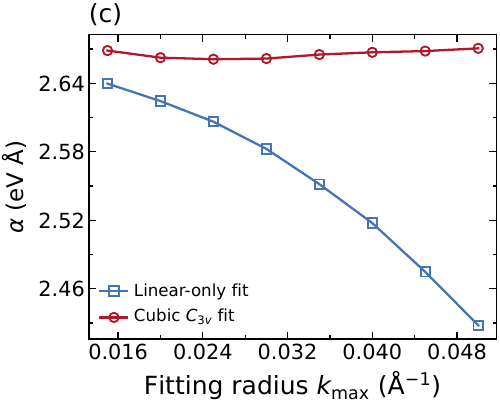}
        \label{fig:panel_c}
    \end{subfigure}
    \hfill
    \begin{subfigure}[t]{0.48\textwidth}
        \centering
     \includegraphics[width=\linewidth]{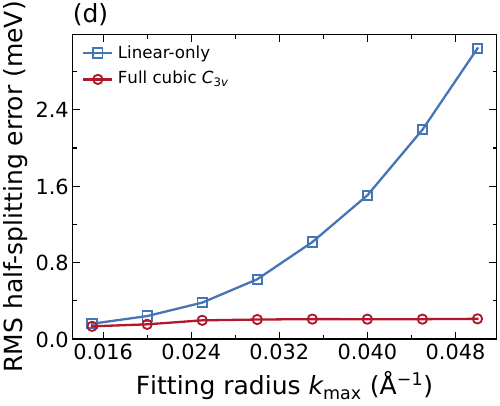}
        \label{fig:panel_d}
    \end{subfigure}
  \caption{Conduction-band fit to the symmetry-constrained $C_{3v}$ Rashba
Hamiltonian. (a),(b) Representative one-dimensional cuts through the fitted
two-dimensional dispersion along $\phi=0^\circ$ and $\phi=30^\circ$,
respectively, for the circular fitting window $k_{\max}=0.050$~\AA$^{-1}$.
Open circles (squares) mark the energy-ordered lower (upper) DFT branch;
solid and dashed maroon lines are the globally fitted two-dimensional
Hamiltonian evaluated along the indicated direction, not fitted
independently to these cuts. In the adopted coordinate convention, the
cubic warping harmonic vanishes along $\phi=0^\circ$ and is maximal along
$\phi=30^\circ$. (c) Fitted linear Rashba coefficient $\alpha_{\mathrm{CB}}$ and
(d) root-mean-square error of the fitted half-splitting, both as a
function of the circular fitting radius $k_{\max}$, comparing a purely
linear Rashba fit (blue squares) with the full third-order $C_{3v}$ model
of Eq.~\ref{eq:kp_hamiltonian} (maroon circles). The plateau in (c) gives the intrinsic
coefficient $\alpha_{\mathrm{CB}}=2.666\pm0.005$~eV\,\AA. The
corresponding drop in RMS error in (d) shows that the cubic correction is
required by the data, not merely permitted by it.}
    
    \label{fig:DFT_data_vs_model_fitting}
\end{figure*}

\subsubsection{Fitting procedure}

At each momentum point, the two DFT bands belonging to a given doublet 
were ordered by energy,
\begin{equation}
E^{\mathrm{DFT}}_{D,-}(\mathbf{k})
<
E^{\mathrm{DFT}}_{D,+}(\mathbf{k}).
\end{equation}
They were then decomposed into a band-center energy,
\begin{equation}
\overline{E}_{D,\mathrm{DFT}}(\mathbf{k})
=
\frac{
E^{\mathrm{DFT}}_{D,+}(\mathbf{k})
+
E^{\mathrm{DFT}}_{D,-}(\mathbf{k})
}{2},
\end{equation}
and a half-splitting,
\begin{equation}
d_{D,\mathrm{DFT}}(\mathbf{k})
=
\frac{
E^{\mathrm{DFT}}_{D,+}(\mathbf{k})
-
E^{\mathrm{DFT}}_{D,-}(\mathbf{k})
}{2}.
\label{eq:half_splitting}
\end{equation}
The Rashba parameters $\alpha_D$, $\beta_D$, and $|\lambda_D|$ were 
obtained by fitting $d_{D,\mathrm{DFT}}(\mathbf{k})$ directly to 
Eq.~(\ref{eq:model_half_splitting}) using nonlinear least squares. Fitting the 
physical parameters directly is important because an expansion of 
$d_D^2$ contains the dependent combinations $\alpha_D^2$, 
$2\alpha_D\beta_D$, $\beta_D^2$, and $\lambda_D^2$, rather than 
independent linear coefficients.

The fits were performed over circular momentum regions
\begin{equation}
k\leq k_{\max},
\end{equation}
centered at $\Gamma$. The analysis was repeated for several values of 
$k_{\max}$ to determine whether the extracted $k\rightarrow0$ coefficient 
depends on the selected momentum range. Equal total weight was assigned to 
each nonempty radial annulus, preventing the larger number of Cartesian-grid 
points at greater $k$ from dominating the fit.

Because $\lambda_D$ enters the energy splitting through $\lambda_D^2$, 
the energy fit determines only its magnitude $|\lambda_D|$. In addition, 
the energy splitting is invariant under the simultaneous transformation
\begin{equation}
(\alpha_D,\beta_D)
\longrightarrow
(-\alpha_D,-\beta_D).
\end{equation}
The overall sign convention for the in-plane coefficients was therefore 
fixed using the helicity of the calculated physical-spin texture. The 
spin texture was also used to verify the expected angular sign pattern of 
the warping term. The numerical coefficients themselves were obtained 
from the energy splitting because the model pseudospin need not coincide 
quantitatively with the physical-spin operator.

\subsubsection{Conduction-band doublet}
\label{sec:cb_fit}

For the conduction-band doublet, the fitting-window dependence was examined over
\begin{equation}
0.015
\leq
k_{\max}
\leq
0.050\,\mathrm{\text{\AA}}^{-1}.
\label{eq:cb_fit_windows}
\end{equation}
The fitted linear coefficient remains between
$2.661$ and $2.670~\mathrm{eV\,\text{\AA}}$ throughout this interval. Taking the window-to-window variation as the dominant systematic uncertainty gives
\begin{equation}
\alpha_{\mathrm{CB}}
=
2.666\pm0.005~\mathrm{eV\,\text{\AA}}.
\label{eq:alpha_cb}
\end{equation}
The weak dependence on $k_{\max}$ shows that this value represents a well-defined intrinsic $k\rightarrow0$ coefficient rather than a slope determined by one particular fitting window.

The isotropic cubic coefficient is negative,
\begin{equation}
\beta_{\mathrm{CB}}
\simeq
-140\pm10~\mathrm{eV\,\text{\AA}^3}.
\label{eq:beta_cb}
\end{equation}
Consequently, the effective in-plane Rashba coefficient decreases with increasing momentum:
\begin{equation}
\alpha_{\mathrm{CB}}^{\mathrm{eff}}(k)
=
\alpha_{\mathrm{CB}}
+
\beta_{\mathrm{CB}}k^2.
\label{eq:alpha_eff_cb}
\end{equation}
For the representative fit at
$k_{\max}=0.050~\mathrm{\text{\AA}^{-1}}$, the parameters
$\alpha_{\mathrm{CB}}=2.669~\mathrm{eV\,\text{\AA}}$ and
$\beta_{\mathrm{CB}}=-146.1~\mathrm{eV\,\text{\AA}^3}$ give
\begin{equation}
\alpha_{\mathrm{CB}}^{\mathrm{eff}}(0.05)
\simeq
2.30~\mathrm{eV\,\text{\AA}}.
\label{eq:alpha_eff_cb_outer}
\end{equation}
Thus, the cubic radial term reduces the effective in-plane coupling at the edge of the fitting region by approximately $14\%$ relative to its $k\rightarrow0$ value.

This correction is also evident when the full model is compared with a purely linear Rashba fit. Over the same
$k_{\max}=0.050~\mathrm{\text{\AA}^{-1}}$ region, the linear-only fit gives an apparent coefficient of
\begin{equation}
\alpha_{\mathrm{CB}}^{\mathrm{linear}}
=
2.428~\mathrm{eV\,\text{\AA}},
\label{eq:alpha_cb_linear}
\end{equation}
whereas the cubic model recovers an intercept of approximately
$2.670~\mathrm{eV\,\text{\AA}}$. The root-mean-square error in the half-splitting decreases from
$3.04~\mathrm{meV}$ for the linear model to
$0.21~\mathrm{meV}$ for the complete cubic model. The negative cubic radial correction is therefore quantitatively required even within the relatively small near-$\Gamma$ region considered here.

The warping coefficient is less tightly constrained because its contribution to the squared splitting scales as
$\lambda_{\mathrm{CB}}^2k^6$. It becomes resolvable only in the wider fitting windows. We therefore quote
\begin{equation}
|\lambda_{\mathrm{CB}}|
\sim
90~\mathrm{eV\,\text{\AA}^3}
\label{eq:lambda_cb}
\end{equation}
as a finite-window estimate rather than as a high-precision intrinsic parameter.

Figure~\ref{fig:DFT_data_vs_model_fitting}(a) and \ref{fig:DFT_data_vs_model_fitting}(b) show one-dimensional cuts through the fitted two-dimensional dispersion along
$\phi=0^\circ$ and $\phi=30^\circ$, respectively. These curves are not independent fits. All parameters were obtained from the complete circular region
$k\leq k_{\max}$, after which the fitted Hamiltonian was evaluated along the two selected directions. Along
$\phi=0^\circ$, the warping harmonic vanishes, and the comparison mainly tests the isotropic radial correction. Along
$\phi=30^\circ$, the warping contribution is maximal. The agreement along both directions therefore tests the radial and angular parts of the same two-dimensional fit.

\subsubsection{Valence-band doublet}
\label{sec:vb_fit}

The same symmetry-constrained Hamiltonian was applied independently to the valence-band-edge doublet. Since the valence splitting is much smaller than the conduction-band splitting, a denser momentum grid was used in a smaller region around $\Gamma$. The fitting-window dependence was examined over
\begin{equation}
0.008
\leq
k_{\max}
\leq
0.030~\mathrm{\text{\AA}^{-1}}.
\label{eq:vb_fit_windows}
\end{equation}

For the smallest windows, the cubic contributions are comparable to the sub-meV numerical residuals. The simultaneous fit of $\alpha_{\mathrm{VB}}$, $\beta_{\mathrm{VB}}$, and $\lambda_{\mathrm{VB}}$ consequently becomes strongly correlated, and the cubic coefficients vary rapidly with the fitting radius.

A comparatively stable linear coefficient is obtained over the intermediate range
\begin{equation}
0.018
\leq
k_{\max}
\leq
0.030~\mathrm{\text{\AA}^{-1}}.
\label{eq:vb_stable_windows}
\end{equation}
Within this interval, the fitted values of
$\alpha_{\mathrm{VB}}$ lie between
$0.342$ and $0.348~\mathrm{eV\,\text{\AA}}$. Taking their window-to-window variation as the systematic uncertainty gives
\begin{equation}
\alpha_{\mathrm{VB}}
=
0.345\pm0.005~\mathrm{eV\,\text{\AA}}.
\label{eq:alpha_vb}
\end{equation}
This value represents the intrinsic near-$\Gamma$ linear coefficient.

The valence-band fits indicate a positive isotropic cubic correction. Its magnitude, however, decreases from approximately
$57$ to $35~\mathrm{eV\,\text{\AA}^3}$ as
$k_{\max}$ increases from
$0.018$ to $0.030~\mathrm{\text{\AA}^{-1}}$. We therefore do not assign a precision value to
$\beta_{\mathrm{VB}}$. The warping coefficient
$\lambda_{\mathrm{VB}}$ is even less well constrained: it is driven close to zero in several smaller windows and becomes resolvable only in the largest window. No quantitative value of
$\lambda_{\mathrm{VB}}$ is consequently reported.

For comparison, a purely linear fit over the broader window
$k_{\max}=0.050~\mathrm{\text{\AA}^{-1}}$ gives
\begin{equation}
\alpha_{\mathrm{VB}}^{\mathrm{linear}}
=
0.382~\mathrm{eV\,\text{\AA}}.
\label{eq:alpha_vb_effective}
\end{equation}
This larger value is a finite-window effective slope rather than the intrinsic $k\rightarrow0$ coefficient. The difference between
Eqs.~(\ref{eq:alpha_vb}) and
(\ref{eq:alpha_vb_effective}) is consistent with the positive higher-order radial correction to the valence-band splitting.

The final fitting results are summarised in
Table~\ref{tab:kp_parameters}. The conduction-band linear and cubic coefficients are well resolved, whereas the weaker valence-band splitting permits a reliable extraction only of the intrinsic linear coefficient.

\begin{table}[t]
\caption{
Summary of the Rashba parameters extracted from the independent conduction- and valence-band two-band fits. The quoted conduction-band coefficients are obtained from the complete fitting-window analysis. For the valence-band doublet, only the intrinsic linear coefficient is treated as quantitatively resolved. The linear-only values at
$k_{\max}=0.050~\mathrm{\text{\AA}^{-1}}$ are finite-window effective slopes and are listed for comparison.
}
\label{tab:kp_parameters}
\centering
\begin{ruledtabular}
\begin{tabular}{lccc}
Doublet and model
&
$\alpha$ or $\alpha^{\mathrm{linear}}$
&
$\beta$
&
$|\lambda|$
\\
&
$(\mathrm{eV\,\text{\AA}})$
&
$(\mathrm{eV\,\text{\AA}^3})$
&
$(\mathrm{eV\,\text{\AA}^3})$
\\
\hline
CB, full $C_{3v}$ model
&
$2.666\pm0.005$
&
$-140\pm10$
&
$\sim90$
\\
CB, linear only
&
$2.428$
&
---
&
---
\\
VB, full $C_{3v}$ model
&
$0.345\pm0.005$
&
not converged
&
not resolved
\\
VB, linear only
&
$0.382$
&
---
&
---
\end{tabular}
\end{ruledtabular}
\end{table}

\section{Topological properties}
\label{sec:topology}

Previous experiments have established the BSTS family as a topological-insulator platform. Here we verify explicitly that the same ordered polar structure used for the Rashba analysis remains a strong topological insulator.

\subsection{Wannier representation}

\begin{figure}[t]
    \centering
    \includegraphics[width=0.95\columnwidth]{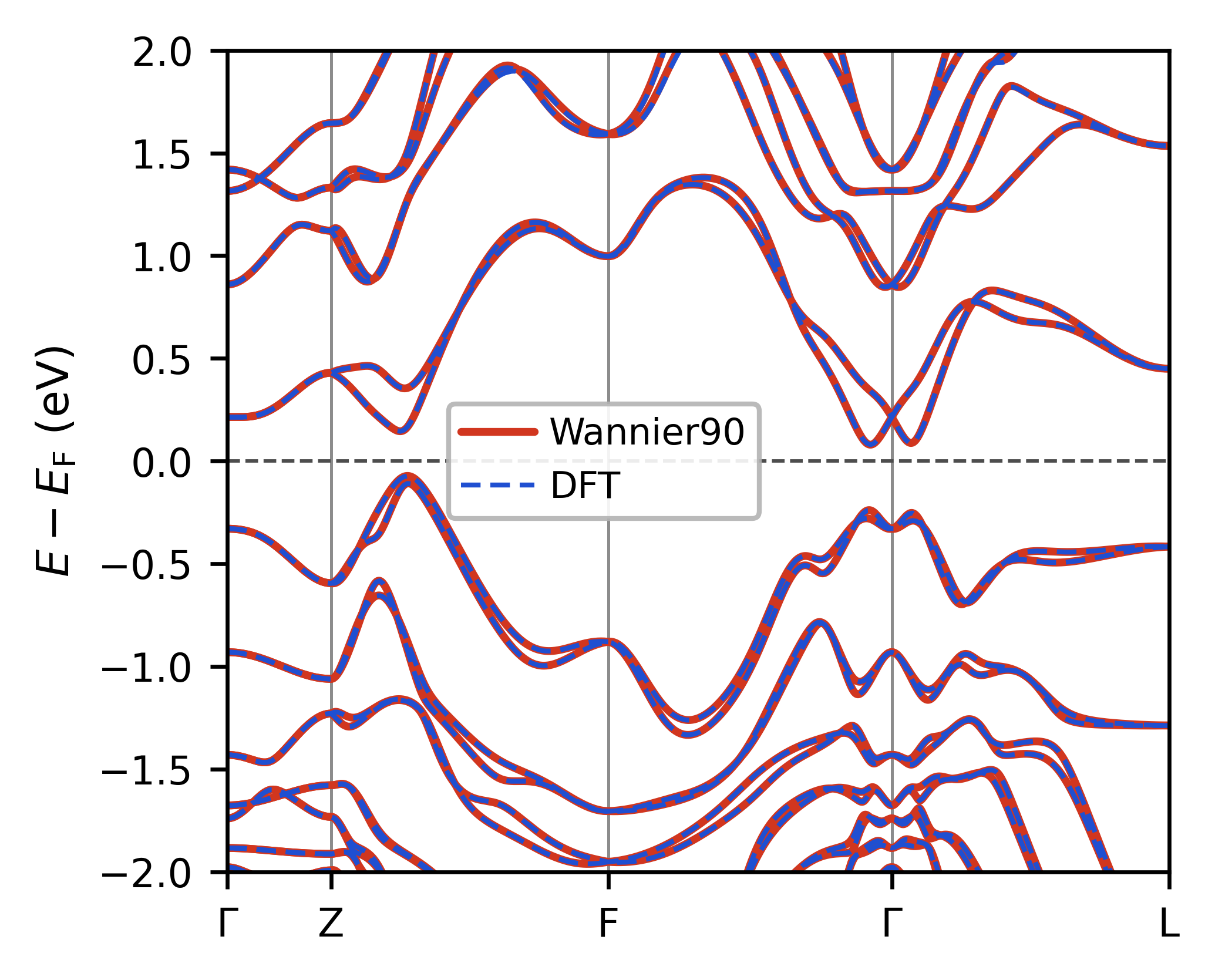}
    \caption{
    Comparison between the first-principles DFT band structure and the
    Wannier90-interpolated bands. The good agreement near the Fermi level
    confirms that the spinor Wannier Hamiltonian accurately captures the
    inverted band manifold and is suitable for evaluating the topological
    invariant and surface spectral function.
    }
    \label{fig:wannier_dft_comparison}
\end{figure}

The low-energy band inversion is dominated by the $p$-orbital manifold of Bi, Sb, Te, and Se. Projection of the SOC Kohn--Sham states onto these orbitals produces a 30-band spinor Wannier Hamiltonian. Figure~\ref{fig:wannier_dft_comparison} compares the Wannier-interpolated bands with the first-principles bands. Agreement in the occupied manifold, around the inverted gap, and throughout the Rashba-split band-edge region validates the Hamiltonian for the subsequent Wilson-loop and surface calculations.

\begin{figure}[t]
    \centering
    \includegraphics[width=0.95\columnwidth]{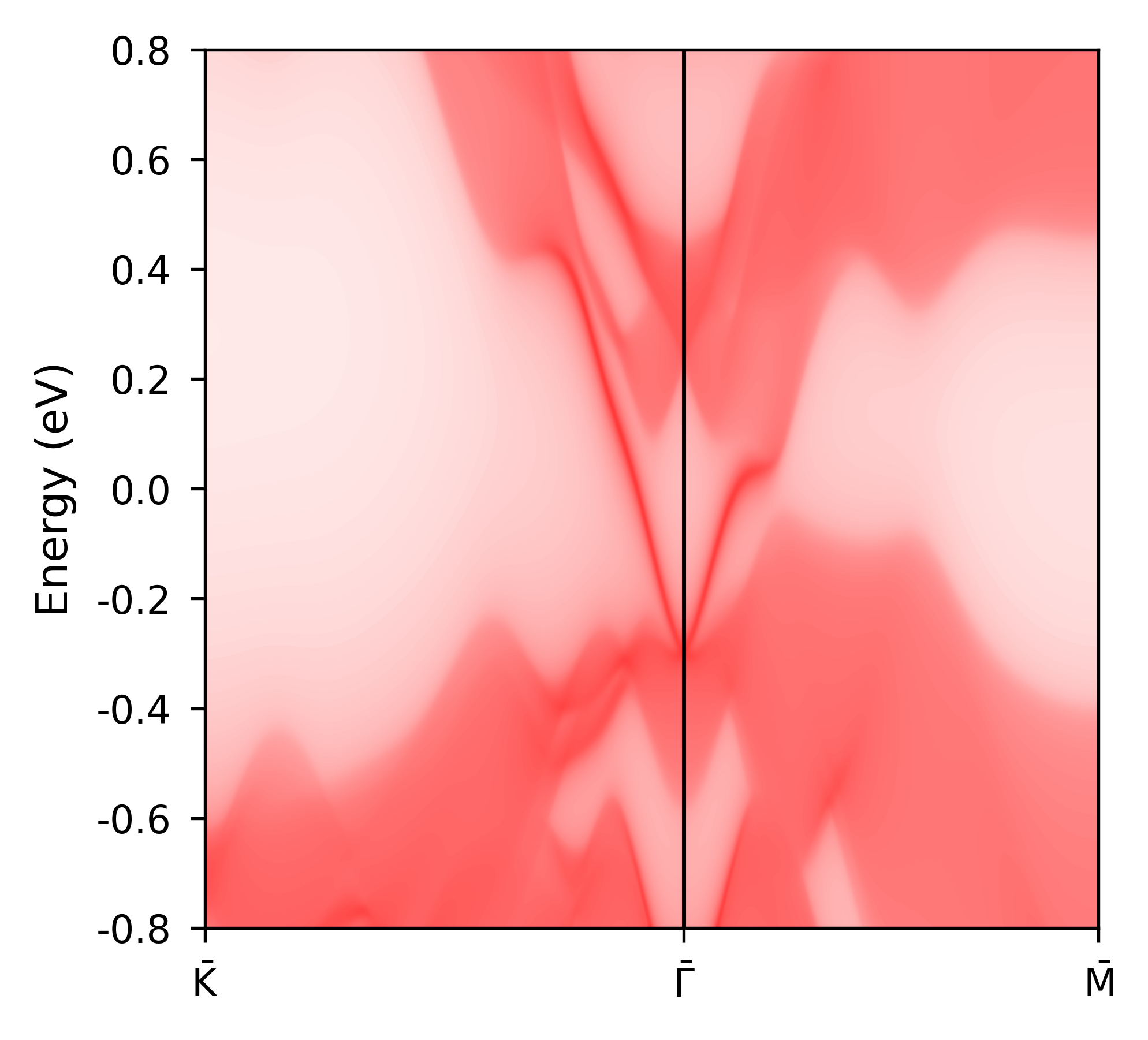}
\caption{Surface spectral function of semi-infinite BiSbTeSe$_2$ in the ordered 
$R3m$ phase for the Te-terminated surface, calculated from the Wannier 
tight-binding Hamiltonian along the surface high-symmetry path 
$\bar{K}$--$\bar{\Gamma}$--$\bar{M}$. The color intensity represents 
the surface-projected spectral weight in arbitrary units, with darker 
regions indicating larger spectral weight. The linearly dispersing 
surface state around $\bar{\Gamma}$ forms a Dirac-like cone connecting 
the projected bulk valence and conduction manifolds, consistent with the 
strong-topological-insulator phase.}
    \label{fig:Te-surface_state}
\end{figure}

\subsection{$\mathbb{Z}_2$ invariant from Wilson loops}

Because the $R3m$ structure lacks inversion symmetry, the topological invariant cannot be inferred from parity eigenvalues at the time-reversal-invariant momenta. We instead evaluate the Wilson-loop evolution of the occupied hybrid Wannier charge centers. 
The resulting indices have strong component $\nu_0=1$, confirming the strong topological phase.

\subsection{Termination-dependent surface spectral function}
\label{sec:surface_spectral_function}

The surface spectral functions were calculated from the Wannier 
tight-binding Hamiltonian using the iterative Green-function method for 
semi-infinite surfaces. The surface plane is spanned by $\mathbf{a}_1$ 
and $\mathbf{a}_2$ and is therefore perpendicular to the crystallographic 
$C_3$ axis. Owing to the polar Se--Bi--Se--Sb--Te stacking sequence, 
the two opposite surfaces are chemically inequivalent. In the 
\textsc{WannierTools} convention used here, the right surface is 
Te-terminated, whereas the left surface is Se-terminated.

Figure~\ref{fig:Te-surface_state} shows the surface spectral function of the Te-terminated surface 
along the $\bar{K}$--$\bar{\Gamma}$--$\bar{M}$ path; the corresponding 
result for the Se-terminated surface is shown in Fig.~\ref{fig:S_surface_Se}. Both terminations 
exhibit a sharply defined, approximately linearly dispersing surface state 
centered at $\bar{\Gamma}$, resembling the characteristic Dirac-like 
surface dispersion of the tetradymite topological insulators 
Bi$_2$Se$_3$ and Bi$_2$Te$_3$~\cite{Zhang2009, zhang2012surface}. 
The energy of the crossing, however, depends strongly on the surface 
termination.

For the Te-terminated surface, the crossing lies approximately 
$0.30$~eV below the Fermi level, similar to that of Bi$_2$Te$_3$, 
whereas for the Se-terminated surface it occurs much closer to the Fermi 
level, at a binding energy of approximately $0.10$~eV, similar to that of 
Bi$_2$Se$_3$. The termination-dependent shift originates from the 
inequivalent surface potentials produced by the chemically distinct 
topmost and bottommost layers of the quintuple layer. Surface potentials 
can modify the Dirac-point energy and the detailed surface dispersion 
without removing the surface state, provided that the protecting 
symmetries remain intact. In contrast to ideal centrosymmetric 
Bi$_2$Se$_3$ and Bi$_2$Te$_3$, whose opposite natural cleavage surfaces 
are chemically equivalent, ordered BiSbTeSe$_2$ therefore provides an 
additional surface degree of freedom. Selection of the terminating layer can shift the
surface-state crossing relative to the chemical potential and may
thereby offer greater flexibility for controlling surface carrier
density, spin--charge conversion, and the relative contributions of
bulk and surface transport channels.


\section{Conclusion}
\label{sec:conclusion}

We have shown that chemically ordered BiSbTeSe$_2$ in the Se--Bi--Se--Sb--Te stacking sequence realizes a polar, noncentrosymmetric topological-insulator phase. The loss of inversion symmetry reduces the symmetry at $\Gamma$ from $D_{3d}$ to $C_{3v}$ and allows a linear-in-$k$ spin splitting of the bulk bands. Fully relativistic first-principles calculations reveal Rashba-split conduction- and valence-band doublets with predominantly tangential in-plane spin textures and a weaker threefold modulation of the spin component parallel to the crystallographic $C_3$ axis.

The bulk splitting was quantified using independent symmetry-constrained two-band $k\cdot p$ models for the conduction- and valence-band-edge Kramers doublets. For the conduction-band doublet, the fitted linear coefficient remains stable over the range
$0.015\leq k_{\max}\leq0.050~\mathrm{\text{\AA}^{-1}}$, yielding
$\alpha_{\mathrm{CB}}
=
2.666\pm0.005~\mathrm{eV\,\text{\AA}}$.
The dispersion also requires a sizeable negative isotropic cubic correction,
$\beta_{\mathrm{CB}}
\simeq
-140\pm10~\mathrm{eV\,\text{\AA}^3}$,
which reduces the effective Rashba coupling at finite momentum. The cubic warping coefficient is less precisely constrained and is estimated to be
$|\lambda_{\mathrm{CB}}|\sim90~\mathrm{eV\,\text{\AA}^3}$.
For the more weakly split valence-band doublet, the near-$\Gamma$ analysis gives
$\alpha_{\mathrm{VB}}
=
0.345\pm0.005~\mathrm{eV\,\text{\AA}}$,
whereas the corresponding cubic coefficients cannot be extracted with comparable reliability from the present data.

The same ordered structure remains topologically nontrivial. The spinor Wannier Hamiltonian reproduces the first-principles bands around the inverted gap and the Rashba-split band edges, while the Wilson-loop evolution and surface spectral function establish the presence of a topological surface state. The chemically inequivalent Te- and Se-terminated
surfaces also place the Dirac-point crossing at different
energies, providing an additional means of tuning the
relative contributions of bulk and surface carriers. Ordered BiSbTeSe$_2$ therefore provides a material platform in which strong bulk spin--momentum locking coexists with a topological surface channel.

The coexistence of bulk-Rashba states and a topological surface channel makes ordered BiSbTeSe$_2$ a promising platform for spin--charge conversion and current-induced spin polarization. Future work should quantify the Edelstein response, spin-transport anisotropy, and the relative contributions of bulk and surface states as the chemical potential is varied. Spin-resolved photoemission and transport measurements could directly test the predicted helicities and large conduction-band splitting. It will also be important to determine how sublattice disorder, gating, and strain modify the Rashba coupling and its spintronic functionality.


\begin{acknowledgments}
The author thanks the Jawaharlal Nehru Centre for Advanced Scientific Research (JNCASR) for financial support through a postdoctoral fellowship. Computational resources were provided by the Param Yukti supercomputer under the National Supercomputing Mission (NSM) Computational Facility. The author also thanks Prof. Awadhesh Narayan (Indian Institute of Science, Bangalore) and Dr. Arijit Sinha (Universit\'e de Li\`ege, Belgium) for insightful discussions.
\end{acknowledgments}

\section*{Data Availability}
The data supporting the findings of this study, including the relaxed crystal structures, Wannier Hamiltonians, fitted model parameters and the scripts used for data analysis are available from the author upon reasonable request.

\bibliographystyle{prb-unsrt}   
\bibliography{refer-new}

\newpage

\setcounter{figure}{0}         
\renewcommand{\thefigure}{S\arabic{figure}} 




\section*{Supplemental Material}

\subsection*{Crystal Structure and Band Inversion}

\begin{figure}[!ht]
    \centering
    \includegraphics[width=0.95\columnwidth]{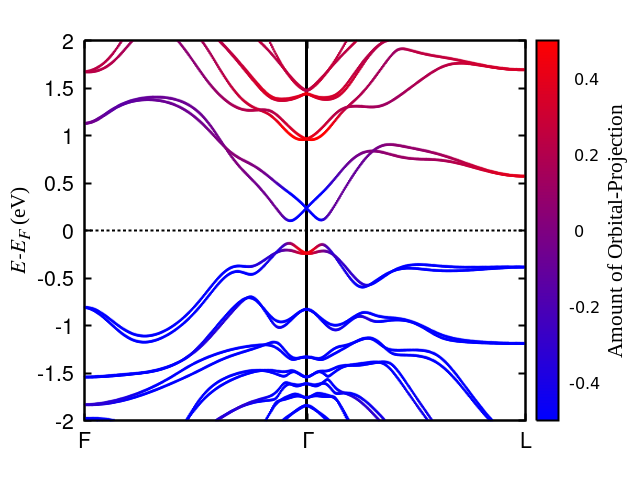}
    \caption{
    Band structure of the primitive cell calculated using DFT with spin-orbit coupling (SOC). The bands are projected onto cation (Bi-6$p$ and Sb-5$p$) and anion (Se-4$p$ and Te-5$p$) orbitals. The color bar indicates relative contributions: +1 denotes 100\% cationic and $-1$ denotes 100\% anionic character. A clear band inversion between the cationic and anionic $p$-orbitals is observed at the $\Gamma$ point.}
    \label{fig:S1}
\end{figure}

We considered three different possible structures for BSTS. They differed in the order of stacking of planes within a quintuple layer: (1) Se-Bi-Se-Sb-Te, (2) Se-Sb-Se-Bi-Te, and (3) Se-Sb-Bi-Te-Se. Of these, stacking sequence (1) was found to lead to an equilibrium structure that was lowest in energy, and this is the structure that was used to compute the band structure presented in Fig.2 of main manuscript and Fig.~\ref{fig:S1}. 

If one uses this oblique rhombohedral primitive unit cell (see Fig.~1(b) of the main manuscript), the lattice parameters are obtained as $|\mathbf{a}_1| = 4.18 $ \AA, $|\mathbf{a}_2| = 4.18$ \AA, $|\mathbf{a}_3| = 9.87   $ \AA  and angles between them are $\alpha = 102.22\,\SI{}{\degree}$, $\beta = 77.78\,\SI{}{\degree}$   and $\gamma = 120.00\,\SI{}{\degree}$.

Fig.~\ref{fig:S1} shows the corresponding band structure with the incorporation of spin-orbit coupling (SOC). We find a band gap at the zone center $\Gamma$ of 0.54 eV, and an overall band gap of 0.24 eV; these values are likely to be an underestimate, because of the use of the GGA in our calculations. Further, we have projected the bands plotted in Fig.~\ref{fig:S1} onto atomic orbitals, so as to determine the relative contributions from cations (Bi-6$p$ and Sb-5$p$) and anions (Te-5$p$ and Se-4$p$). The color of a band in this figure indicates the relative contributions from anions and cations, according to the color bar shown on the right; a value of $-1$ indicates 100\% anionic contribution, and a value of +1 indicates 100\% cationic contribution. In the neighborhood of the $\Gamma$ point, we see clear evidence of a band inversion: the valence band is red (cationic) whereas the rest of the band is blue (anionic), and the reverse is true for the conduction band. This band inversion is a strong indication that the material is likely to be a topological insulator. 

\subsection*{Calculation of the momentum-resolved bulk spin texture}
\label{sec:si_spin_texture}

The momentum-resolved spin textures were calculated directly from the
fully relativistic spinor wavefunctions obtained using Quantum
ESPRESSO version 7.2. The general post-processing procedure follows
the approach described by Ghosh \textit{et al.},\cite{ghosh2023rashba}
with the momentum sampling adapted here to a two-dimensional region
surrounding the $\Gamma$ point.

A self-consistent calculation including spin-orbit coupling was first
performed using an $18\times18\times9$ Monkhorst--Pack $k$-point mesh.
The resulting converged charge density was subsequently used in a
non-self-consistent calculation on a dense local momentum grid centred
at $\Gamma$.

The momentum grid was constructed in the physical plane perpendicular
to the crystallographic $C_{3}$ axis. We introduced a local
orthonormal Cartesian coordinate system satisfying
\begin{equation}
\hat{\mathbf{z}}\parallel C_{3},
\qquad
\mathbf{k}
=
k_x\hat{\mathbf{x}}+k_y\hat{\mathbf{y}},
\qquad
\mathbf{k}\cdot\hat{\mathbf{z}}=0.
\label{eq:si_spin_plane}
\end{equation}
The momentum window
\begin{equation}
-0.12~\mathrm{\text{\AA}^{-1}}
\leq k_x,k_y
\leq
0.12~\mathrm{\text{\AA}^{-1}}
\label{eq:si_spin_window}
\end{equation}
was divided into a uniform $101\times101$ grid, giving a total of
10\,201 momentum points. All points were included explicitly in the
non-self-consistent calculation. The spin textures were therefore
evaluated directly over the complete local momentum region, without
reconstructing symmetry-related points from the irreducible part of
the Brillouin zone.

Because the primitive unit cell of BiSbTeSe$_2$ is oblique, the
physical plane perpendicular to the crystallographic $C_{3}$ axis
does not correspond, in general, to a plane having a constant third
crystal-coordinate component. The Cartesian momentum points generated
using Eq.~(\ref{eq:si_spin_plane}) were therefore transformed into the
reciprocal crystal coordinates required by Quantum ESPRESSO. As a
result, points lying in the physical $k_x$--$k_y$ plane can have a
nonzero third component when expressed in the reciprocal basis of the
primitive cell.

The $101\times101$ grid exceeds the default number of lines accepted
by the \texttt{bands.x} and \texttt{plotband.x} post-processing
utilities. We therefore modified the corresponding Quantum ESPRESSO
7.2 source files, \texttt{bands.f90} and \texttt{plotband.f90}, before
recompilation. In both files, the parameter controlling the maximum
number of accepted lines was increased to
\begin{equation}
\texttt{max\_lines}=100000.
\end{equation}
This modification allowed the complete set of 10\,201 momentum points
to be processed without truncation. No other part of the Quantum
ESPRESSO source code was modified.

The expectation values of the three spin components were calculated
using the \texttt{bands.x} post-processing program. Separate
calculations were performed by selecting the three values of the
\texttt{lsigma(i)} flag, where $i=1,2,3$ respectively calculates the expectation values
\begin{equation}
\left\langle \sigma_x\right\rangle_{n\mathbf{k}},
\qquad
\left\langle \sigma_y\right\rangle_{n\mathbf{k}},
\qquad
\left\langle \sigma_z\right\rangle_{n\mathbf{k}}
\label{eq:si_sigma_expectation}
\end{equation}
for each spinor eigenstate
$\lvert\psi_{n\mathbf{k}}\rangle$. These quantities are related to
the physical-spin expectation values through
\begin{equation}
\left\langle\mathbf{S}\right\rangle_{n\mathbf{k}}
=
\frac{\hbar}{2}
\left\langle\boldsymbol{\sigma}\right\rangle_{n\mathbf{k}}.
\label{eq:si_physical_spin}
\end{equation}

The output generated by \texttt{bands.x} was subsequently processed
using \texttt{plotband.x}. The resulting spin-expectation-value files
were combined with the Cartesian $k_x$ and $k_y$ coordinates of the
local momentum grid to construct the vector-field data used for the
spin-texture plots.

The four bands closest to the bulk energy gap were extracted at every
momentum point. These correspond to the two uppermost valence-band
branches, denoted VB$-1$ and VB, and the two lowest conduction-band
branches, denoted CB and CB$+1$. At each momentum point, the two
states belonging to a given doublet were ordered according to their
energies. The spin textures of all four energy-ordered branches were
then plotted separately.

In the resulting figures, the arrows represent the spin components
projected onto the plane perpendicular to the $C_{3}$ axis,
\begin{equation}
\mathbf{S}_{\parallel}
=
\left(
\left\langle S_x\right\rangle,
\left\langle S_y\right\rangle
\right),
\end{equation}
whereas the color scale represents the spin component parallel to
the crystallographic $C_{3}$ axis,
\begin{equation}
S_{C_3}
=
\left\langle S_z\right\rangle.
\end{equation}

The two branches of both the conduction-band and valence-band
doublets exhibit opposite spin textures. In particular, the
energy-ordered branches show opposite in-plane helicities at the same
finite momentum, while each individual branch satisfies the
time-reversal relation
\begin{equation}
\left\langle\mathbf{S}\right\rangle_n(-\mathbf{k})
=
-\left\langle\mathbf{S}\right\rangle_n(\mathbf{k}).
\label{eq:si_time_reversal_spin}
\end{equation}
The conduction- and valence-band doublets therefore display the
characteristic spin--momentum locking associated with Rashba
splitting. A weaker threefold angular modulation of the spin component
parallel to $C_{3}$ is also visible and is consistent with the cubic
spin--orbit term allowed by the local $C_{3v}$ symmetry.

\subsection*{Fitting-window dependence of the conduction-band
Rashba parameters}

\begin{figure}[h]
\centering
\includegraphics[width=0.8\linewidth]{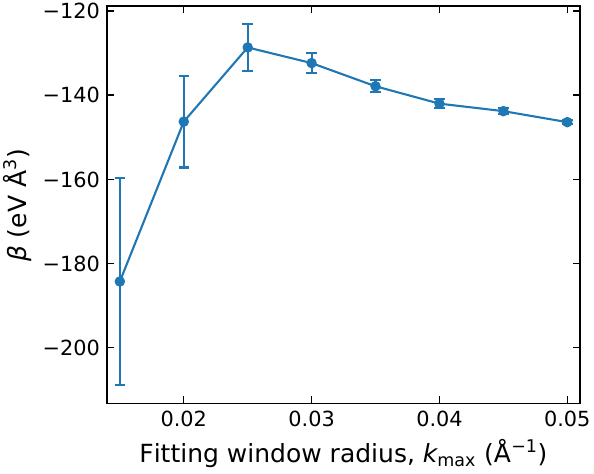}
\caption{Fitted isotropic cubic coefficient $\beta_{\mathrm{CB}}$ as a
function of the fitting radius $k_{\max}$. The coefficient
is less well constrained in the smallest windows, where it
is strongly correlated with $\alpha_{\mathrm{CB}}$ and
$|\lambda_{\mathrm{CB}}|$, and stabilizes at approximately
$-140$ to $-146~\mathrm{eV\,\text{\AA}^3}$ for
$k_{\max}\gtrsim0.030~\mathrm{\text{\AA}^{-1}}$. This behaviour
is consistent with the value
$\beta_{\mathrm{CB}}\simeq-140\pm10~\mathrm{eV\,\text{\AA}^3}$
quoted in Eq.~\ref{eq:beta_cb} of the main text.}
\label{fig:S_beta}
\end{figure}

\begin{figure}[h]
\centering
\includegraphics[width=0.8\linewidth]{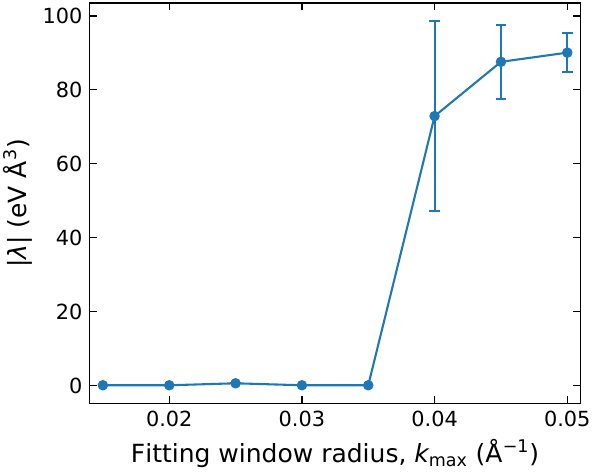}
\caption{Fitted magnitude of the cubic warping coefficient
$|\lambda_{\mathrm{CB}}|$ as a function of the fitting radius
$k_{\max}$. Because its contribution to the squared
half-splitting scales as $\lambda_{\mathrm{CB}}^{2}k^{6}$, the
coefficient is statistically unresolved in the smallest
windows and becomes distinguishable from zero only for
$k_{\max}\gtrsim0.040~\mathrm{\text{\AA}^{-1}}$. At
$k_{\max}=0.050~\mathrm{\text{\AA}^{-1}}$, the fit gives
$|\lambda_{\mathrm{CB}}|=90.1\pm5.2~\mathrm{eV\,\text{\AA}^3}$.
This value should be interpreted as a finite-window
estimate rather than as a precisely determined intrinsic
coefficient.}
\label{fig:S_lambda}
\end{figure}

\begin{figure}[h]
\centering
\includegraphics[width=0.8\linewidth]{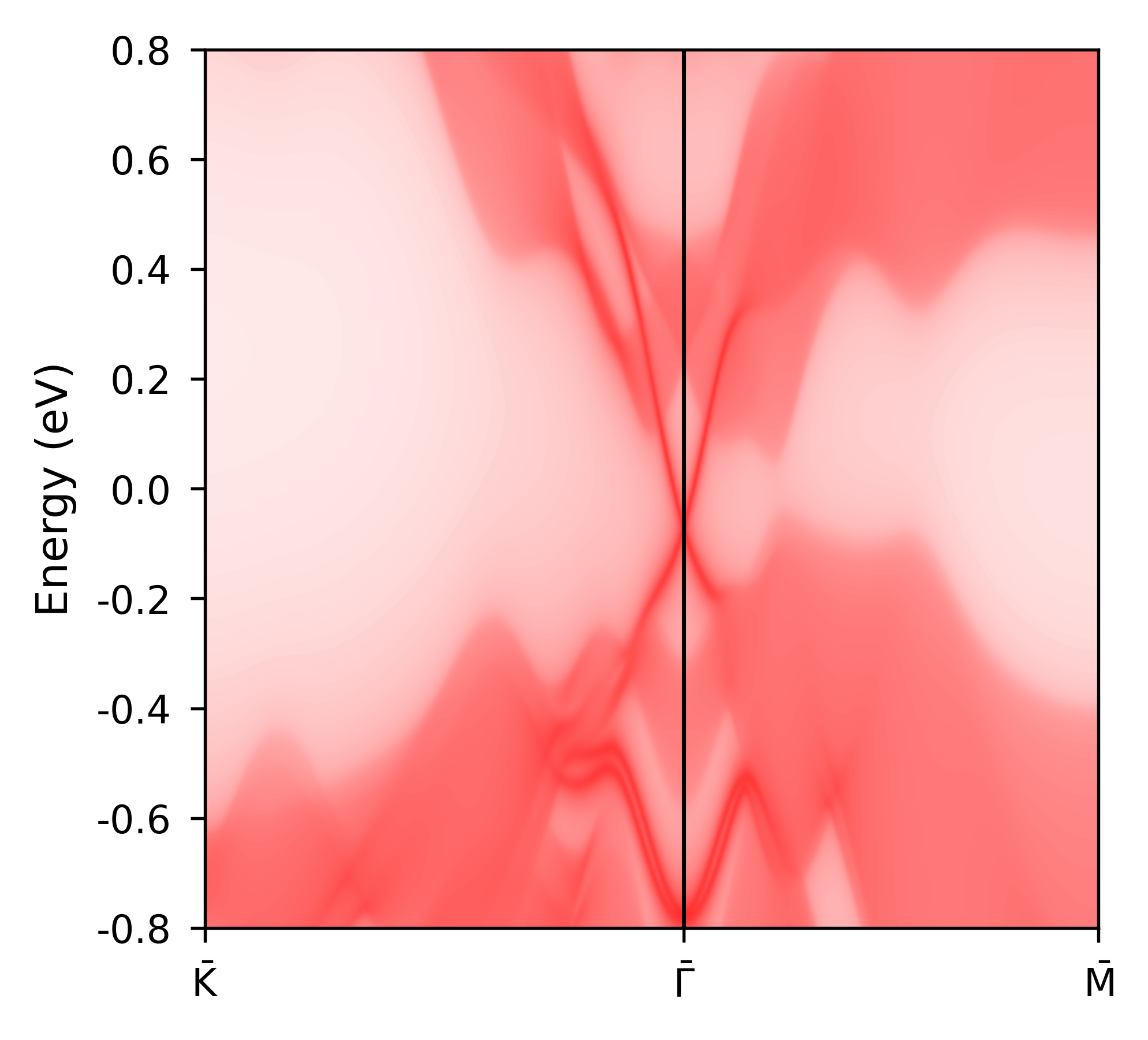}
\caption{
Surface spectral function of semi-infinite
$R3m$-BiSbTeSe$_2$ for the Se-terminated surface,
calculated along the surface high-symmetry path
$\bar{K}$--$\bar{\Gamma}$--$\bar{M}$. The color
intensity represents the surface-projected spectral weight
in arbitrary units. A Dirac-like surface state connects
the projected bulk valence and conduction manifolds, with
the crossing located approximately $0.10$~eV below the
Fermi level.}
\label{fig:S_surface_Se}
\end{figure}

Fig.~\ref{fig:S_beta} and Fig.~\ref{fig:S_lambda} show the fitting-window dependence
of the cubic conduction-band parameters $\beta_{\mathrm{CB}}$
and $|\lambda_{\mathrm{CB}}|$, respectively, over the range
$0.015 \leq k_{\max} \leq 0.050~\text{\AA}^{-1}$ considered
in Sec.~III\,C of the main text.

Unless noted otherwise, error bars are the one-standard-deviation
uncertainties returned by the covariance matrix of the nonlinear
least-squares fit at each $k_{\max}$.

\subsection*{Surface State of Se-terminated Surface}

\end{document}